\documentclass[10pt,journal,compsoc]{IEEEtran}
\usepackage{amsmath,epsfig}
\usepackage{verbatim}
\usepackage{relsize}
\usepackage{lmodern}
\usepackage{amsfonts}
\usepackage{amssymb}
\usepackage{array}
\usepackage{color}
\usepackage{rotating}

\begin{document}\sloppy

\def\x{{\mathbf x}}
\def\L{{\cal L}}

\newcommand{\se}[1]{{\textcolor{black}{#1}}}
\newcommand{\secor}[2]{{\textcolor{black}{#2}}}
\newcommand{\secmt}[1]{{\textcolor{blue}{[#1]}}}
\newcommand{\ie}{\emph{i.e.}}

\newcommand{\ah}[1]{{\textcolor{blue}{#1}}}

\title{EEG-based Inter-Subject Correlation Schemes  in a Stimuli-Shared Framework: Interplay with Valence and Arousal}
%
%

%

\author{\IEEEauthorblockN{
Ayoub Hajlaoui\IEEEauthorrefmark{1}\IEEEauthorrefmark{2},
Mohamed Chetouani\IEEEauthorrefmark{1} 
and Slim Essid\IEEEauthorrefmark{2}}\\
\IEEEauthorblockA{\IEEEauthorrefmark{1} CNRS \& Sorbonne Universit\'es, UPMC Universit\'e Paris 06, Institut des Syst\`emes Intelligents et de Robotique (ISIR) UMR7222, Paris, France}\\
\IEEEauthorblockA{\IEEEauthorrefmark{2} LTCI, T\'el\'ecom ParisTech, Universit\'e Paris--Saclay}}

\maketitle




\begin{abstract}
Affective computing is confronted to high inter-subject variability, in both emotional and physiological responses to a given stimulus. In a stimuli-shared framework, that is to say for different subjects who watch the same \secor{stimulus}{stimuli}, Inter-Subject Correlation (ISC) measured from Electroencephalographic (EEG) recordings characterize the correlations between the respective signals at the different EEG channels. In order to investigate the interplay between ISC and emotion, we propose to study the effect of valence and arousal on the ISC score.  \se{To this end,} we exploited various computational schemes corresponding to different subsets of the dataset: all the data, \secor{stimuli-wise}{stimulus-wise}, subject pairwise, and both \secor{stimuli-wise}{stimulus-wise} and subject pairwise. We \secor{then}{thus} applied these schemes to the HCI MAHNOB and DEAP databases. Our results suggest that the ISC score decreases with valence and increases with arousal, as already shown by previous results on functional MRI. 
%
%
%
%
%
%
%
\end{abstract}
\section{Introduction}
\label{sec:intro}
A significant part of the investigations in affective computing research seeks a better understanding of the link between elicited emotion and physiological responses. More specifically, EEG has attracted the attention of researchers in the field of affective computing and it has been shown to hold precious cues for emotion classification \cite{bajaj2014human,zheng2017multichannel,takahashi2004remarks}. However, affective computing has to \secor{tackle}{cope with} the variability of individual responses to stimuli, whether it be at the emotion level or at the physiological signal level. Indeed, from one subject to another: 
i) the same stimulus can elicit different emotions \secor{from one subject to another}{} \cite{abadi2013multimodal,scherer1994evidence}; and ii) the same elicited emotion translates into different physiological responses \cite{morioka2015learning,koch2007gender}.

Individual differences across subjects limit the generalization of automated emotion classification. Indeed, extracted EEG features vary significantly across individuals \cite{zhu2015cross}, which results in a degradation of classification results when made across subjects \cite{soleymani2012multimodal}. To overcome this issue, researchers can employ subject-specific approaches, \ie~develop computational models using only data from the targeted subject \cite{koelstra2012deap}. Others use different transfer learning strategies to improve EEG-based inter-subject classification \cite{lin2017improving}.

In this paper, we address the inter-subject variation issue from an interaction perspective, adopting a \secor{stimuli-centered}{stimulus-centered} study of synchrony between EEG signals, in the same fashion as the robot-centered approach in robotics \cite{boucenna2014learning}. In other words, we study the correlations between EEG signals of different subjects who watched the same videos, even if they did not watch them simultaneously.

Two main reasons motivate this approach:
\begin{itemize}
 \item Shared experiences, such as the exposure to the same audiovisual content, play an important part in the interactions between individuals.
\item For complex tasks such as \secor{stimuli-based}{stimulus-based} emotion elicitation, single-trial EEG analysis is often a necessity. Therefore, analyzing the signals recorded from different subjects \se{and obtaining insights about their differences and commonalities} can make the results more generalizable. 
\end{itemize}

To simultaneously analyze the EEG signals of different subjects, we use the Inter Subject Correlation (ISC) framework, as described in previous studies \cite{dmochowski2012correlated,dmochowski2014audience,ki2016attention}.  Dependencies between ISC of EEG recorded during audiovisual stimuli and subject conditions such as age or sex have been established. For instance, decrease in ISC of EEG has been shown as ages of the subjects increase \cite{petroni2016age}.

Others have established links between ISC of functional Magnetic Resonance Imaging (MRI) and emotion, showing that ISC increases for specific regions of the brain when the stimulus elicits high arousal or low valence \cite{nummenmaa2012emotions}. Replicating such results with EEG signals would both prove consistency and allow their usability with more lightweight devices.

In line with these previous works, and having acknowledged inter-subject and inter-stimuli variations \cite{abadi2013multimodal}, we propose various schemes to study the effects of valence and arousal variations on ISC of EEG recorded from different subjects watching the same videos: on all the dataset, stimulus-wise, subject-pairwise, or both stimulus-wise and subject-pairwise. Those schemes are detailed in Section \ref{sec:schemes}.

In addition to the establishment of a link between ISC of EEG signals and valence/arousal levels which is, to the best of our knowledge, completely novel, our main contributions are :\begin{itemize}
\item the proposal and comparison of various ISC computational schemes
\item a particular interest given to the statistical validity of the observed ISC variation along valence and arousal dimensions, using computationally intensive randomization tests.
\end{itemize}

Section \ref{sec:principle} is a reminder of the ISC framework. Section \ref{sec:schemes} presents and discusses different ISC computational schemes, whereas Section \ref{sec:emotion} raises the issue of interpretation of ISC results. Sections \ref{sec:hci} and \ref{sec:deap} show the results obtained with different schemes respectively on the HCI MAHNOB \cite{soleymani2012multimodal} and DEAP \cite{koelstra2012deap} databases. Finally, section \ref{sec:limitations} emphasizes some limitations of our work and explains observed differences between the databases.
%
%

\section{The ISC principle}
\label{sec:principle}
\begin{figure}[!h]
\includegraphics[scale=0.35]{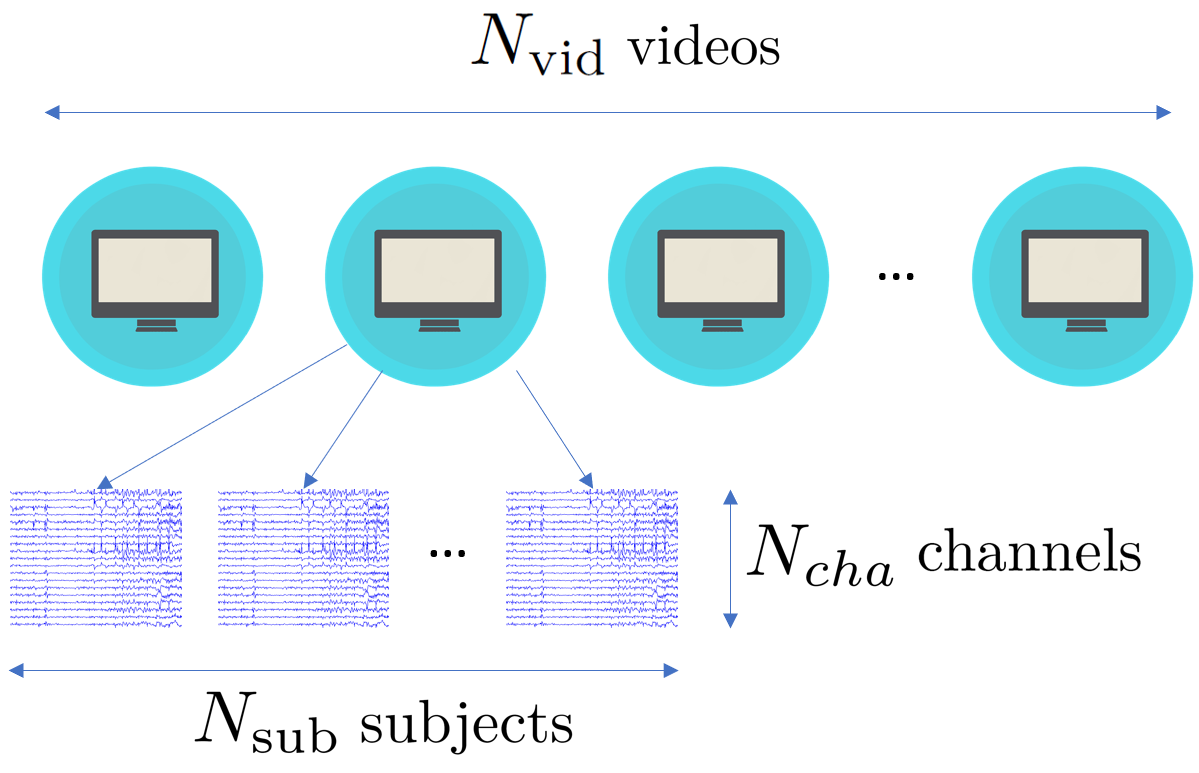}
\caption{Stimulus-centered study of EEG signals}
\label{fig:schema_database}
\end{figure}

\se{To simplify the presentation, we introduce the principle of ISC by directly instantiating it on our use-case:} $N_\text{sub}$ subjects watch $N_\text{vid}$ video stimuli. All subjects watch the same videos. The videos are not watched simultaneously. During each stimulus, EEG signals are recorded from the scalp of each subject with a $N_{cha}$-channel EEG headset. Figure \ref{fig:schema_database} illustrates the situation.

For each video, each subject annotates the emotion felt using the valence and arousal dimensions. The annotation scale can be either discrete or continuous. 
\subsection{ISC score computation}
Let $X_{i,v}$ denote the EEG \se{data} matrix recorded from subject $i$ while he/she was watching video $v$. $i$ ranges from $1$ to $N_\text{sub}$ while $v$ ranges from $1$ to $N_\text{vid}$. $X_{i,v}$ is a $N_\text{cha}\times T_v$ matrix, where $T_v$ is the number of \secor{data points}{EEG signal samples} recorded for each channel, which depends on the length of the video $v$.

Given the matrices $R_{i j}$ of size $N_\text{cha}\times N_\text{cha}$ which each measure the cross-covariance of all electrodes in subject $i$ with all electrodes in subject $j$, the pooled within-subject covariance $R_w$ and the pooled between-subject cross-covariance $R_b$ are defined as follows:
\begin{eqnarray}
R_{i j}&=&\sum_{t=1}^T (X_{i,v}(:,t)-\bar{X}_{i,v})(X_{j,v}(:,t)-\bar{X}_{j,v})' \; ; \\
R_w&=&\dfrac{1}{N_\text{sub}}\sum_{i=1}^{N_\text{sub}}R_{ii} \; ;\\
R_b&=&\dfrac{1}{N_\text{sub}(N_\text{sub}-1)}\sum_{i=1}^{N_\text{sub}}\sum_{j\neq i}R_{ij}.
\end{eqnarray}
where $X'$ denotes the transpose of $X$ and $\bar{X}$ denotes the vector corresponding to the mean over time of $X$. In Section \ref{sec:schemes}, a focus will be made on a pairwise definition of $R_w$ and $R_b$, that is to say pooled over each pair of subjects.  

Given the matrices $R_b$ and $R_w$, the eigenvectors $e_k$ of $R_w^{-1}R_b$ are computed and ranked in decreasing order of associated eigenvalue.  These eigenvectors are then used to compute the correlation strengths $C_k$ in the following fashion:\begin{eqnarray}
C_k = \dfrac{e_k' R_b e_k}{e_k' R_w e_k}.
\end{eqnarray}
$C_k$ accounts for the ratio of the projection strength of $e_k$ on $R_b$ to its projection strength on $R_w$.
Following previous studies that concluded that the choice of the three first components \secor{to be}{is} a good compromise \cite{dmochowski2014audience,petroni2016age,ki2016attention}, we define the ISC score as $C_1+C_2+C_3$.

\subsection{Averaging $R_{ij}$ to compute ISC eigenvectors}
Actually, what is usually done in the EEG-based ISC domain is the averaging of matrices $R_{ij}$ across all stimuli, or across both all stimuli and all pairs of subjects (when ISC are considered pairwise). This only concerns the eigenvectors computation step \cite{petroni2016age}. For instance, when the averaging is done across all stimuli, the averaged matrices $\mathbf{R_{ij}}$ are computed, for each pair of subjects $(i,j)$, in the following manner :
\begin{eqnarray}
\mathbf{R_{ij}} = \dfrac{1}{N_\text{vid}}\sum_{v=1}^{N_\text{vid}} R_{ij}
\end{eqnarray}
Then, following (2) and (3), $R_{b_\text{global}}$ and $R_{w_\text{global}}$ are computed from the averaged matrices $\mathbf{R_{ij}}$.
Eigenvectors $e_k$ are then computed from $R_{w_\text{global}}^{-1}R_{b_\text{global}}$.

\subsection{Shrinkage}
\label{shrinkage}

As proposed in \cite{blankertz2011single} for Linear Discriminant Analysis-based single-trial ERP classification, $R_{w_\text{global}}$ may be shrunk to improve robustness to outliers. \se{Let} $\gamma$ be a regularization parameter between $0$ and $1$ and $\bar{\lambda}$ the mean eigenvalue of $R_{w_\text{global}}$:

\begin{eqnarray}
R_{w_\text{global}} \leftarrow (1-\gamma)R_{w_\text{global}}+\gamma\bar{\lambda}I
\end{eqnarray}
When estimating a big covariance matrix, large eigenvalues are estimated too large, and small eigenvalues are estimated too small \cite{blankertz2011single}. Shrinkage modifies extreme eigenvalues towards the average eigenvalue. What is convenient is that shrinkage does not change the eigenvectors of such covariance matrices. In addition to dampening the effect of outliers by this modification, shrinkage allows to compute the inverse of \se{the} shrunk $R_{w_\text{global}}$ when $R^{-1}_{w_\text{global}}$ cannot be computed. 

\section{Different ISC computational schemes}
\label{sec:schemes}
In this paper, we exploit our \secor{stimuli-shared}{shared stimuli} framework, to define different ISC computational schemes following theses perspectives:
\begin{itemize}
\item whether to compare the EEG signals of the subjects pairwise or globally;
\item how to combine the data on which to compute the eigenvectors of $R_w^{-1}R_b$?: \se{that is whether to consider} all the dataset, stimulus-wise, subject-pairwise, or both stimulus-wise and subject-pairwise \se{data batches}.
\end{itemize}

\subsection{Comparing subject signals globally vs pairwise}
Computing ISC eigenvectors using the signal recordings of all $N_\text{sub}$ subjects globally suits the case when we wish to compare each subject to the group. In this case, ISC scores are computed for each subject $i$ using the following expressions:
\begin{eqnarray}
(C_k)_i &=& \dfrac{e_k' (R_b)_i e_k}{e_k' (R_w)_i e_k}\se{;}\\
\text{where}~~(R_b)_i&=&\dfrac{1}{N-1}\sum_{j\neq i}(R_{ij}+R_{ji})\se{;}\\
\text{and}~~(R_w)_i&=&\dfrac{1}{N-1}\sum_{j\neq i}(R_{ii}+R_{jj})\se{.}
\end{eqnarray}
In our attempt to establish a link between emotion and ISC scores, we could compare, for each video, each subject to the rest, and look at the effect of elicited emotion on the ISC score of each subject. However, doing so would compel us to consider annotation agreement globally, whereas considering annotation agreement pairwise allows a finer distinction between agreement and non-agreement. In the pairwise setting, we compute the ISC score for each pair of subjects $(i,j)$ in the following fashion:
\begin{eqnarray}
(C_k)_{ij} &=& \dfrac{e_k' (R_b)_{i,j} e_k}{e_k' (R_w)_{i,j} e_k}\se{;}\\
\text{where}~~(R_b)_{ij}&=&R_{ij}+R_{ji}\se{;}\\
\text{and}~~(R_w)_{ij}&=&R_{ii}+R_{jj}\se{.}
\end{eqnarray}
%
%

We chose to focus on this pairwise setting. In fact, in addition to allowing one to consider agreement in a pairwise fashion, this multiplies the ISC data on which to study valence and arousal effects.

\subsection{Choosing the data on which to compute the eigenvectors}
\begin{itemize}
\item Averaging the matrices $R_{ij}$ across all stimuli, and then computing the eigenvectors $e_k$ from $R_{w_\text{global}}^{-1}R_{b_\text{global}}$, that is using the whole dataset (all subjects, all stimuli), generalizes such eigenvectors and makes them more robust to outliers. All the available information is used to compute the covariance matrices, thus allowing \se{a} better precision. In that fashion, we seek to maximize inter-subject correlation on all the dataset. We refer to this scheme as \boldmath$V_\text{all}$\unboldmath. However, as EEG responses are very subject-dependent and session-dependent, computing the eigenvectors $e_k$ on more specific subsets can also be considered.

\item Rather than being computed from $R_{w_\text{global}}^{-1}R_{b_\text{global}}$, the eigenvectors $e_k$  can be computed stimulus-wise, that that is separately for each stimulus, on all pairs of subjects, therefore taking stimulus-dependency into account. The assumption is that we wish to maximize ISC for each stimulus separately. Practically, it consists in not averaging matrices $R_{ij}$ on all stimuli, but rather in processing each stimulus separately.

This scheme, presented in Figure \ref{fig:schema_stimuluswise}, is referred to as \boldmath$V_\text{stim}$\unboldmath.
\begin{figure}[!h]
\begin{center}
\includegraphics[scale=0.25]{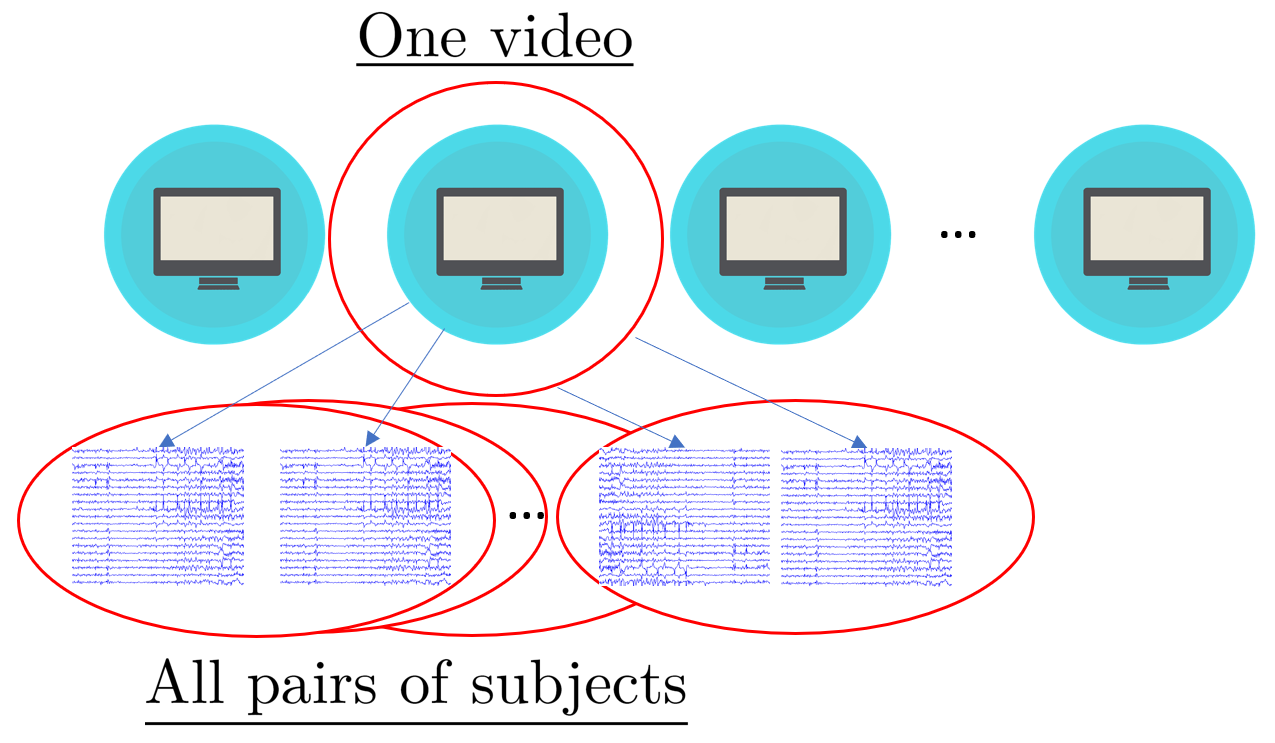}
\caption{Data on which the eigenvectors $e_k$ are computed in the case of \boldmath$V_\text{stim}$\unboldmath}
\label{fig:schema_stimuluswise}
\end{center}
\end{figure}

\item The eigenvectors $e_k$ can also be computed subject-pairwise, that is separately for each pair of subjects, on all stimuli, as shown in Figure \ref{fig:schema_subjectwise}. Thus, subject-dependency is taken into account. Mathematically, for subjects $i$ and $j$, this means that the sums in equations (2) and (3) are respectively replaced by $(R_b)_{ij}$ and $(R_w)_{ij}$ (equations (10) and (11)). We refer to this scheme as \boldmath$V_\text{pair}$\unboldmath.
\begin{figure}[!h]
\begin{center}
\includegraphics[scale=0.25]{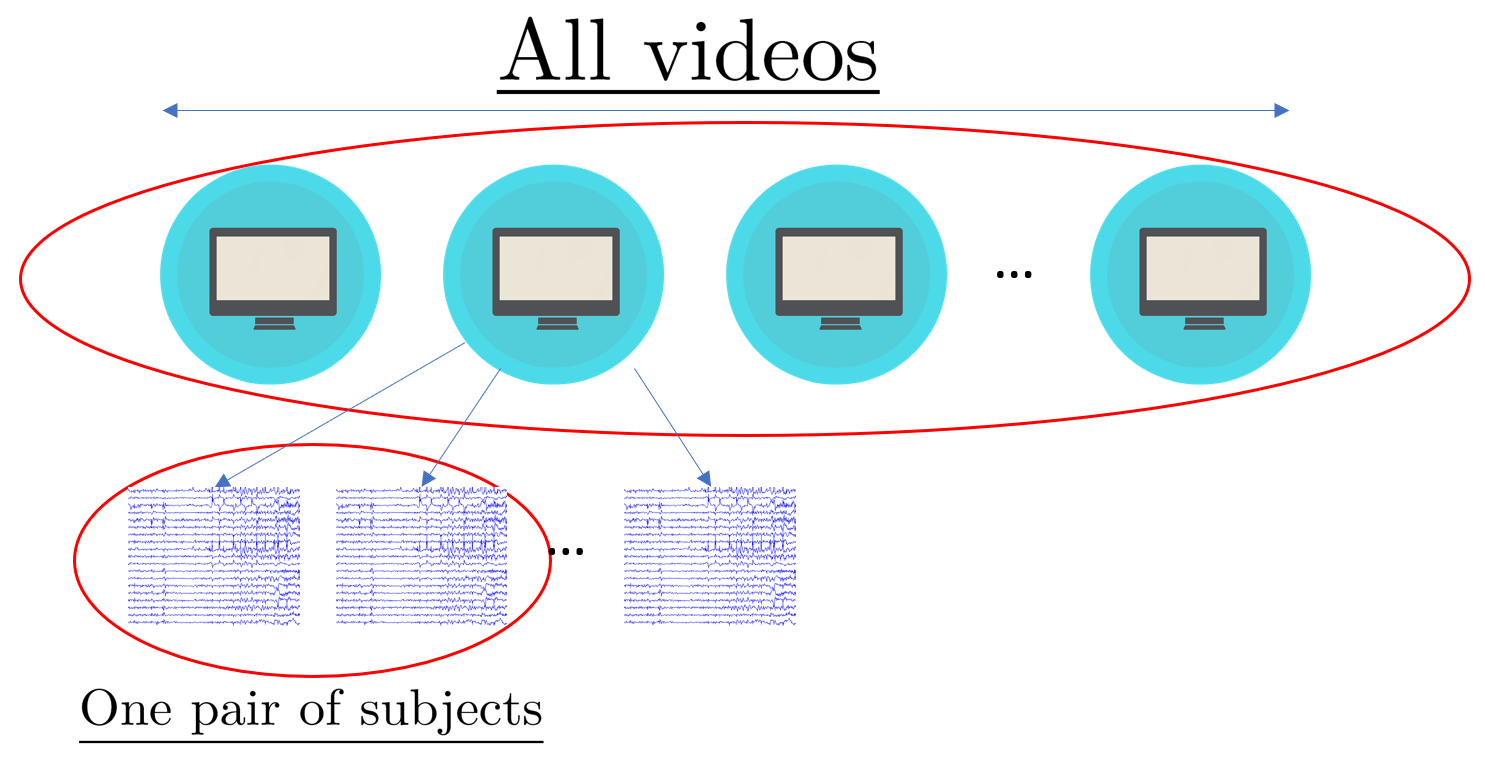}
\caption{Data on which the eigenvectors of $R_{w_\text{global}}^{-1}R_{b_\text{global}}$  are computed in the case of \boldmath$V_\text{pair}$\unboldmath}
\label{fig:schema_subjectwise}
\end{center}
\end{figure}

\item Finally, the eigenvectors $e_k$ can be computed both stimulus-wise and subject-pairwise. This takes both specificities into account, which seems well suited for EEG analysis. However, in this way, covariance matrices are estimated on smaller portions of the dataset, which automatically induces a drop in precision in the estimation of those covariance matrices. We refer to this scheme as \boldmath$V_\text{stim/pair}$\unboldmath.
\end{itemize}
\section{Studying effects of emotion on ISC}
\label{sec:emotion}
There are $N_\text{pairs}=\dfrac{N_\text{sub}(N_\text{sub}-1)}{2}$ pairs of subjects. Regardless of the slicing scheme (Section \ref{sec:schemes}-B), $N_\text{pairs}$ associated ISC scores are obtained for each video, which makes a total of $N_\text{pairs}\times N_\text{vid}$ ISC scores. For each pair of subjects, one has to take a decision regarding their agreement on the valence or the arousal annotations, respectively. Indeed, to establish a link between the emotion experienced by two subjects and the ISC score between their EEG signals, we limit the study to the cases where the subjects agree on the annotation of the emotion.

Then, pairs of subjects for which there is agreement should be classified according to the level of valence or arousal that was annotated.

In the HCI MAHNOB database \se{used in this study (see Section~\ref{sec:hci})}, valence and arousal annotations are discrete values in $\{1,2,...,9\}$. We divide valence and arousal annotations in $3$ classes : $\{1,2,3\}$ are considered low, $\{4,5,6\}$ are considered average, and $\{7,8,9\}$ are considered high, following the usual division made in the literature, and more specifically in the paper introducing HCI MAHNOB.

In the DEAP database \se{(see Section~\ref{sec:deap})}, valence and arousal annotations are continuous values in $[1;9]$. We \secor{also}{again} divide valence and arousal annotations in $3$ classes: values in $[1;3.5]$ are considered low, values in $]3.5;6.5[$ are considered average, and values in $[6.5;9]$ are considered high.

\subsection{Assessing pairwise agreement}
\label{assessing_pairwise}
Assessing the agreement of each pair of subjects is a difficult task that may first seem arbitrary. Previous works have used the Cohen's kappa score as an agreement indicator \cite{litman2003recognizing}. However, as this score is suited to multi-annotator cases, its use is less interesting when only computed on a given pair of subjects, which is our case. In addition, we do not wish to assess the agreement of each pair of subjects on all videos, but rather on each video. Therefore, our focus is on the assessment of agreement both subject-pairwise and stimulus-wise. We introduce ad hoc rules for such an assessment, taking into account the non-linearity of agreement \cite{martinez2014don}:

\begin{itemize}
\item For a given stimulus, we assume that two annotations from the same category (low, average, high, as previously defined) are in agreement with each other.

\item We consider two annotations from different categories to be in agreement with each other if and only if their difference is lower or equal to 1.
\end{itemize}

Such rules are chosen both to correspond to the usual categories in the literature (low, average, high) and to allow for some agreement flexibility at the border between two classes.

\begin{figure}[!h]
\includegraphics[scale=0.35]{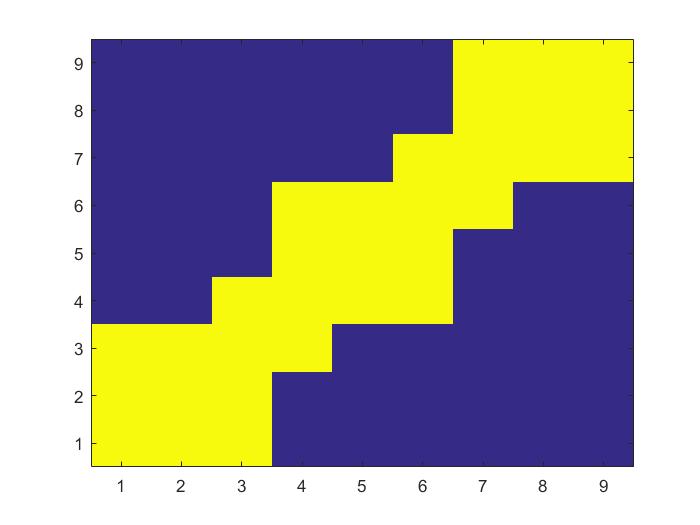}
\caption{Agreement decision matrix (axis values represent annotations from both subjects; yellow stands for agreement)}
\label{fig:agreement}
\end{figure}

Figure \ref{fig:agreement} sums up those rules in the form of a decision matrix for the HCI MAHNOB case. For instance, for a given video stimulus, if subject $i$ annotates a valence of $2$ and subject $j$ a valence of $4$, they are considered in disagreement with each other. On the contrary, if subject $i$ gives an annotation of $7$ and subject $j$ an annotation of $9$, their annotations are considered to agree with each other.

\subsection{Assigning a subject pairwise annotation for a given stimulus when there is agreement}
When two subjects agree on the annotation of a given stimulus, we want to assign a common label to this video, which is specific to this pair of subjects, in order to establish a link between this label and the ISC score. Previous works use majority decisions to assign a global annotation to each stimulus \cite{aigrain2016multimodal}. However, this is not relevant when only considering two annotators, nor is it justified when the annotations are not binary.

Therefore, for a given stimulus and a given pair of subjects who agree on the annotation of this stimulus, we decide to assign the mean of their two annotations as the pair annotation of this stimulus.
\subsection{Effects of valence and arousal on ISC}
\label{statistics}
For each category of annotation (low, average, high), the mean ISC of all pairs of subjects who agree on the annotation and whose pairwise mean annotation is in this category is computed, to establish a link between the annotation category and the mean ISC score of this category. To do so, the significance of the difference between \secor{each category's mean}{the mean ISC scores of different categories} has to be assessed. Usually, parametric tests such as t-tests or ANOVA procedures are performed. Even if transformations---such as Fisher's transforms before a t-test---can be applied to make the data better fit the assumptions of the tests, these assumptions are still unwarranted.

Other approaches consist in the comparison of the empirically obtained ISC scores to simulated ISC scores on surrogates of the data. The inconvenient is that for statistical validity to hold, the computation of ISC scores from scratch has to be repeated an important number of times.

Rather, our approach is inspired from the randomization test proposed in \cite{yeh2000more}. Given the ISC scores \secor{spread out}{separately computed} in the 3 valence (or arousal) categories, we shuffle these ISC scores $2^{20}$ times, reassigning each score randomly to one of the 3 categories (each category's cardinal being kept constant). To assess the significance of the difference between the mean \se{ISC scores} obtained for two categories,  we look at the number $n$ of the $2^{20}$ shuffles that gave a higher difference of means than the one experimentally obtained. The significance level of the real ISC difference obtained between the two categories is at most $\dfrac{n+1}{2^{20}+1}$\cite{noreen1989computer}. This non-parametric test allows us to assess the significance of our results without the need of complex unwarranted hypotheses on ISC score distributions. With this significance test, we are able to assess whether the variations on ISC that we observe as a function of assessed emotion are significant or not.

This procedure is performed to compare ISC scores from different valence \secor{and}{or} arousal categories, thus trying to assess the dependencies between the valence (resp. arousal) level and the ISC score.

Let us note that significance values not only depend on differences of means, but also on the cardinal of each category, which explains how a slight difference can be more significant than a larger one.

\section{Results on HCI MAHNOB}
\label{sec:hci}
HCI MAHNOB \cite{soleymani2012multimodal} is a multi-modal dataset where various physiological signals were recorded from subjects who watched video stimuli. Among these physiological recordings, we are interested in the EEG signals.

Each subject assessed the emotion elicited by each stimulus in terms of valence and arousal. With our notations, $N_\text{vid}=20$ and $N_\text{sub}=24$ (we only took into account the subjects who watched all the videos). This gives a total of $5520$ pairwise ISC scores, among which $3685$ agreements on valence, and $2968$ agreements on arousal. Following \ref{assessing_pairwise}, we restrict our computations on pairs of subjects where agreement is obtained.

The focus is made on two specific schemes, that are \boldmath$V_\text{all}$\unboldmath~and \boldmath$V_\text{stim/pair}$\unboldmath. The two remaining schemes are discussed \se{more briefly}. Significance results correspond to the upper bounds obtained with the method presented in \ref{statistics}.

\subsection{Results with \boldmath$V_\text{all}$\unboldmath}
\begin{figure}[!h]
\begin{center}
\includegraphics[scale=0.35]{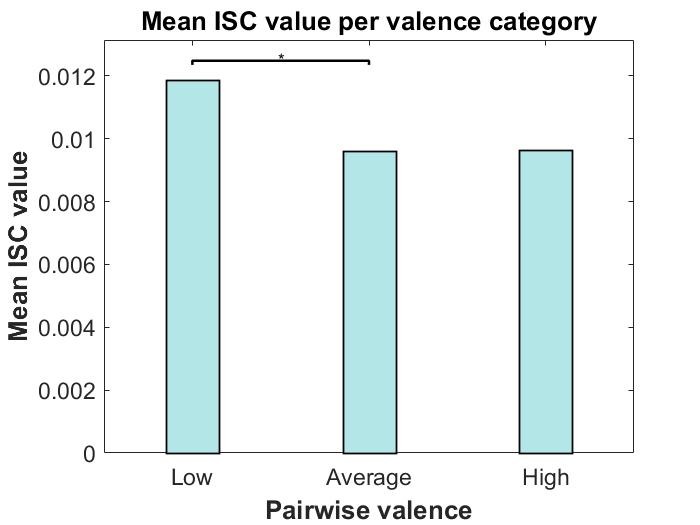}
\caption{Mean ISC score per valence category (low, average, high) for \boldmath$V_\text{all}$\unboldmath~~
*,**,***: significance at the respective levels of 5\%, 1\%, and 0.1\% (HCI MAHNOB database)}
\label{fig:val1}
\end{center}
\end{figure}
\begin{figure}[!h]
\includegraphics[scale=0.35]{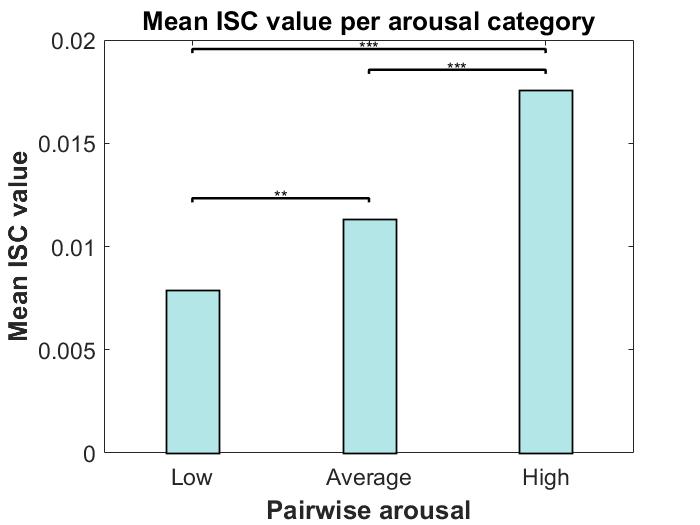}
\caption{Mean ISC score per arousal category (\boldmath$V_\text{all}$\unboldmath, HCI MAHNOB)}
\label{fig:aro1}
\end{figure}
Figures \ref{fig:val1} and \ref{fig:aro1} show the means of pairwise ISC scores for each category of annotation (low, average, and high), respectively for valence and arousal, along with information on the significance of the difference between each category. The considered significance levels are $5\%$, $1\%$ and $0.1\%$.

As shown in Figure \ref{fig:val1}, ISC scores obtained in this fashion decrease when valence increases. In other words, low valence elicitation induces better Inter Subject Correlation, which echoes the findings of Nummenmaa et al. \cite{nummenmaa2012emotions}, the latter restricting such variation to specific regions of the brain. However, only the difference between low valence ISC scores and average valence ISC scores is significant at the 5\% level.

As for the arousal dimension, Figure \ref{fig:aro1} reveals an increase of ISC scores when arousal increases, which was also expected. In terms of significance, such raise is easier to observe than the decrease of ISC along valence.
%
%

\subsection{Results with \boldmath$V_\text{stim/pair}$\unboldmath}
\begin{figure}[!h]
\includegraphics[scale=0.35]{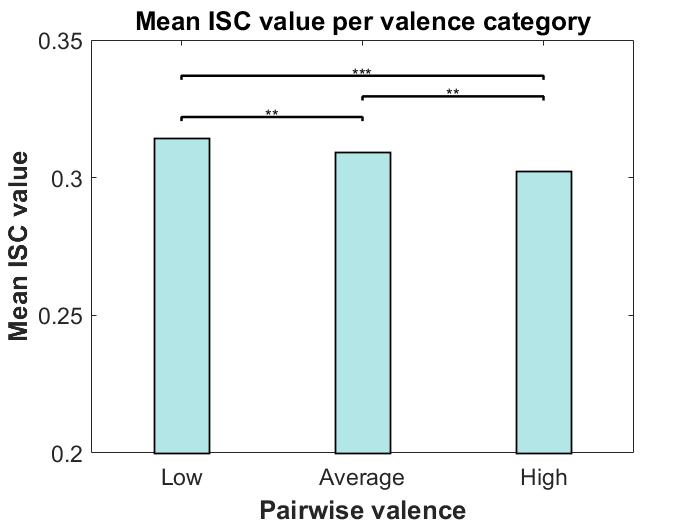}
\caption{Mean ISC score per valence category (\boldmath$V_\text{stim/pair}$\unboldmath)}
\label{fig:val4}
\end{figure}
\begin{figure}[!h]
\includegraphics[scale=0.35]{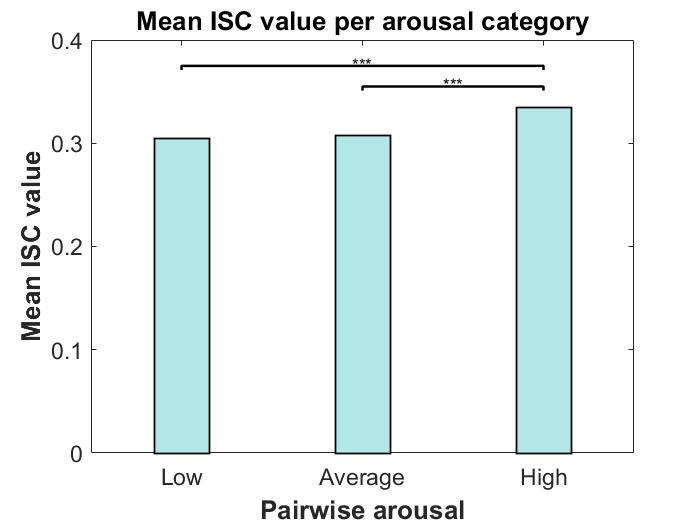}
\caption{Mean ISC score per arousal category (\boldmath$V_\text{stim/pair}$\unboldmath, HCI MAHNOB)}
\label{fig:aro4}
\end{figure}

Contrary to \boldmath$V_\text{all}$\unboldmath, this scheme takes into account both subject pair dependency and stimulus dependency. Let us see how the obtained results back the previous ones, despite this dependency change.

Figure \ref{fig:val4} shows the same tendency as Figure \ref{fig:val1} in terms of ISC decrease when valence increases. However, differences are better in term of significance. Figure \ref{fig:aro4} also shows the same tendency as Figure \ref{fig:val1}, but the significance level between low arousal ISC and average arousal ISC is decreased.

The monotonicity of ISC as a function of valence and a function of arousal is strengthened as it is observed for both schemes. In addition, one can notice that computing ISC eigenvectors separately for each pair of subjects and each stimulus yields more significant results for valence, whereas it degrades significance for arousal. This could be interpreted by a \secor{less}{lesser} subject and stimulus dependency of arousal. The following subsection suggests a difference between valence and arousal annotations that could explain the phenomenon.

\subsection{Linking the ISC level to the annotation agreement}

It is worth noticing that among the 5520 HCI MAHNOB data points on which ISC can be computed (276 subject pairs $\times$ 20 video stimuli) :\\
- 3685 correspond to a pairwise valence annotation agreement whereas the remaining 1835 correspond to a pairwise valence annotation disagreement (using the definitions presented in Section \ref{sec:emotion}); \\
- 2968 correspond to a pairwise arousal annotation agreement whereas the remaining 2552 correspond to a pairwise arousal annotation disagreement.

At first glance, one could conclude that agreement occurs more easily on valence than on arousal. However, it is more interesting to go in depth with a comparison of ISC levels according to valence (respectively arousal) agreement/disagreement. The results of such \se{a} comparison are \secor{made}{given} in Table \ref{table:agreement} (ISC scores were computed using the scheme \boldmath$V_\text{all}$\unboldmath, HCI MAHNOB).
\begin{table}[!h]
\centering
\caption{Comparison of mean ISC scores obtained in case of annotation agreement/disagreement}
\begin{tabular}{|c|c|c|c|} \hline
Dimension&Agreement&Disagreement&Significance\\ \hline
Valence&0.0104&0.0106&0.46\\ \hline
Arousal&0.0112&0.0097&0.052\\ \hline
\end{tabular}
\label{table:agreement}
\end{table}

Table \ref{table:agreement} shows that the mean ISC score is higher on the data subset where agreement on arousal occurs than on the one where there is disagreement on arousal annotation. Such \se{a} difference is almost significant at the 5 \% level. As for valence annotation, there is almost no ISC difference between agreement and disagreement cases.

This could mean that even if its occurs less frequently, agreement on arousal is more consistent than agreement on valence. Further, it could explain why the ISC monotonicity as a function of valence is more significant when ISC eigenvectors are computed separately for each pair of subject and each stimulus, rather than on the whole dataset.

\section{Results on DEAP}
\label{sec:deap}
DEAP \cite{koelstra2012deap} is another multi-modal dataset where various physiological signals, among which EEG signals, were recorded from subjects. The main difference with HCI MAHNOB is that the emotions were elicited by the means of music video stimuli. With our notations, $N_\text{vid}=40$ and $N_\text{sub}=32$. This gives a total of $19840$ pairwise ISC scores, among which $11126$ agreements on valence, and $9184$ agreements on arousal.
\subsection{Results with \boldmath$V_\text{all}$\unboldmath}
\begin{figure}[!h]
\begin{center}
\includegraphics[scale=0.35]{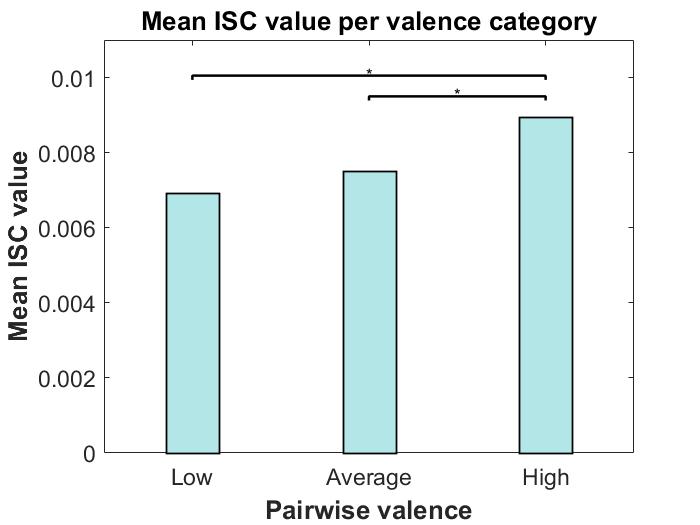}
\caption{Mean ISC score per valence category ( \boldmath$V_\text{all}$\unboldmath, DEAP)}
\label{fig:val1deap}
\end{center}
\end{figure}

Figure \ref{fig:val1deap} shows that contrary to HCI MAHNOB, mean ISC scores increase when valence increases, even if the significance is only at the level of 5\%. Reasons why such \se{a} difference is observed are discussed in \ref{differences_hci_deap}.

As for the arousal dimension, Figure \ref{fig:aro1} reveals a variation similar to the one obtained for HCI MAHNOB, that is to say an increase of ISC scores when arousal increases, only with a less satisfying significance.

\begin{figure}[!h]
\begin{center}
\includegraphics[scale=0.35]{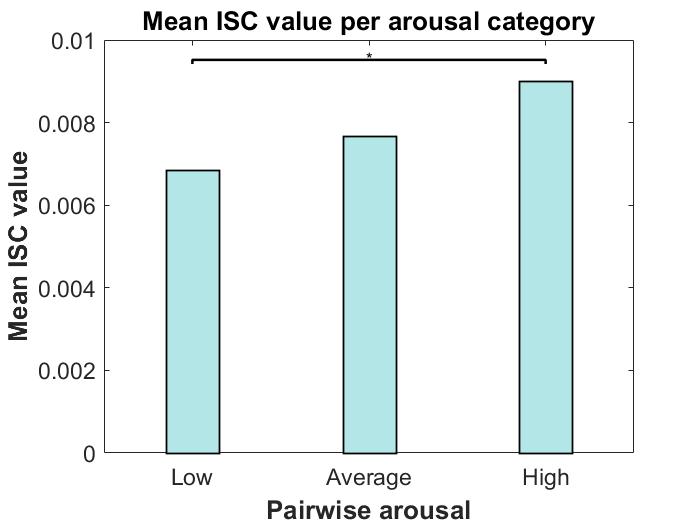}
\caption{Mean ISC score per arousal category ( \boldmath$V_\text{all}$\unboldmath)}
\label{fig:aro1deap}
\end{center}
\end{figure}
\subsection{Results with \boldmath$V_\text{stim/pair}$\unboldmath}
\begin{figure}[!h]
\begin{center}
\includegraphics[scale=0.35]{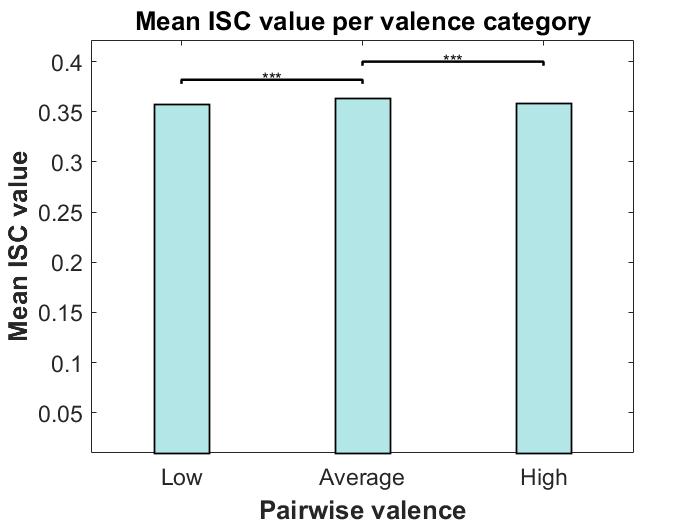}
\caption{Mean ISC score per valence category ( \boldmath$V_\text{stim/pair}$\unboldmath, DEAP)}
\label{fig:val4deap}
\end{center}
\end{figure}

\begin{figure}[!h]
\begin{center}
\includegraphics[scale=0.35]{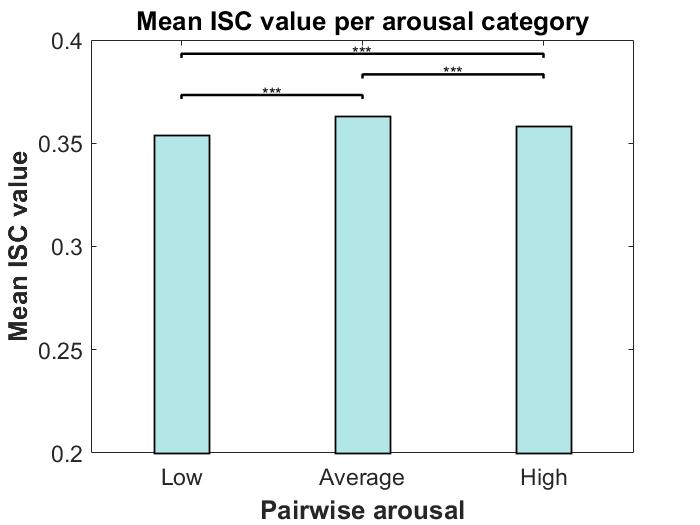}
\caption{Mean ISC score per arousal category ( \boldmath$V_\text{stim/pair}$\unboldmath, DEAP)}
\label{fig:aro4deap}
\end{center}
\end{figure}

When ISC eigenvectors are computed subject-pairwise and stimulus-wise, a different pattern of variations is observed for both valence (Figure \ref{fig:val4deap}) and arousal (Figure \ref{fig:aro4deap}). Indeed, there is a significant ISC decrease for extreme values of valence or arousal. The mean ISC obtained for average valence (resp. arousal) is higher.

However, we can notice something quite consistent with the results concerning HCI MAHNOB, that is to say a significant decrease in ISC between low and high valence, and a significant increase in ISC between low and high arousal. 

\section{Further discussion}
\label{sec:limitations}

\subsection{Agreement is arbitrarily defined}
The assessment of subject-pairwise agreement introduced in Section \ref{sec:emotion} follows arbitrary rules, even \secor{if the latter}{though they} were carefully chosen for consistency. Performing a calibration phase before presenting the stimuli to each participant could help homogenizing the meaning of annotation values among subjects, and therefore mitigate this arbitrary aspect.
\subsection{ISC score variation from one scheme to another}
Comparing ISC score levels obtained from the different schemes, one can clearly notice that the more specific the slicing scheme (Section \ref{sec:schemes}-B), the higher the ISC scores. This is quite natural as the correlation is maximized on smaller, more specific subsets of the data.

\subsection{Differences of ISC score variations along valence between HCI MAHNOB and DEAP}
\label{differences_hci_deap}
In the case of HCI MAHNOB, the ISC score clearly decreases along the valence dimension (Figures \ref{fig:val1} and \ref{fig:val4}). However, results are more mitigated in the case of DEAP (Figures \ref{fig:val1deap} and \ref{fig:val4deap}). This can be explained by both the different nature of \se{the} stimuli \se{used} and the annotation procedure. Annotation is continuous in DEAP, whereas it is discrete in HCI MAHNOB.

But some more striking comparison between HCI MAHNOB and DEAP annotation results could explain this difference better. Table \ref{table:difference1} shows that the mean absolute valence annotation difference is significantly higher for DEAP than for HCI MAHNOB. Significance is computed using the method described in \ref{statistics}. One could wonder if the difference observed is simply due to the annotation nature, which is discrete in the case of HCI MAHNOB and continuous for DEAP. However, the same comparison for arousal yields a smaller difference between the two databases, even if the difference is still significant. Therefore, Table \ref{table:difference1} shows a difference between the databases that could explain why the ISC score clearly decreases along the valence dimension in the case of HCI MAHNOB, whereas it is more mitigated in the case of DEAP.
\begin{table}[!h]
\centering
\caption{Mean absolute value of pairwise valence annotation difference}
\begin{tabular}{|c|c|c|} \hline
Dimension&Valence&Arousal\\ \hline
HCI MAHNOB&1.49&2.02\\ \hline
DEAP&1.69&2.10\\ \hline
HCI/DEAP difference significance&$<10^{-5}$&$2.5\times 10^{-4}$\\ \hline
\end{tabular}
\label{table:difference1}
\end{table}

\secor{After that comparison made on the whole databases}{Further}, it is interesting to compare the same quantities between HCI MAHNOB and DEAP with a restriction to the agreement cases, using the definitions of agreement exposed in \ref{assessing_pairwise}. This is relevant as the ISC scores we presented were computed on agreeing pairs of subjects. Such \se{a} comparison is made in Table \ref{table:difference2}. Again, \secor{it}{this} shows that overall, the agreement level is significantly better in the case of HCI MAHNOB than DEAP, with a more significant difference for the valence dimension. This would support the hypothesis that the different valence agreement levels between the two databases explain the difference between ISC variations along valence.

\begin{table}[!h]
\centering
\caption{Mean absolute of pairwise valence annotation difference among cases of agreement}
\begin{tabular}{|c|c|c|} \hline
Dimension&Valence&Arousal\\ \hline
HCI MAHNOB&0.77&0.84\\ \hline
DEAP&0.80&0.86\\ \hline
HCI/DEAP difference significance&$0.0057$&$0.05$\\ \hline
\end{tabular}
\label{table:difference2}
\end{table}
\subsection{Effects of shrinkage}

As exposed in \ref{shrinkage}, $R_{w_\text{global}}$ may be shrunk to improve robustness to outliers, by the means of a regularization parameter $\gamma$  between $0$ and $1$. This regularization parameter has a limited effect on significance but practically none on the variation itself.
\section{Conclusions and future work}
\label{sec:conclusions}
We have presented and described various schemes to study the effects of valence and arousal on EEG Inter Subject Correlation between participants who watched the same audiovisual stimuli. We have introduced a definition of agreement so as to limit our study on agreeing subject pairs. Finally, we have presented the obtained results for two schemes on the HCI MAHNOB and DEAP affective datasets \cite{soleymani2012multimodal,koelstra2012deap}.

Our results show a consistent  increase in ISC scores when arousal increases. Along the valence dimension, a consistent decrease in ISC was obtained in the case of HCI MAHNOB, whereas this conclusion is more mitigated for DEAP. The different nature of \secor{DEAP stimuli}{the stimuli used in the DEAP dataset} (music videos) can explain such drawbacks, as well as the difference between discrete/continuous annotations and, more importantly, the finer agreement level in HCI MAHNOB.

Both the decrease in ISC scores when valence increases and the increase in ISC scores when arousal increases are consistent with previous results on functional MRI in the literature \cite{nummenmaa2012emotions}.

\secor{Particular attention was given}{A great deal of attention was devoted} to the significance of such variations, using computationally intensive randomization tests. \secor{What is interesting}{Of particular note} is the fact these results are backed by the different schemes. Even if each scheme focuses on a different dependency (stimuli-wise, subject pairwise...), there is a clear trend when it comes to the variation of ISC score as a function of valence or arousal.
%
%


In future work, the study will be extended to datasets where a calibration phase is available, and more focus will be made on the definition of pairwise annotation agreement.

\section*{Acknowledgments}

This research has been supported by the Laboratory of
Excellence SMART (ANR-11-LABX-65) supported by French
State funds managed by the ANR within the Investissements
d'Avenir programme (ANR-11-IDEX-0004-02).


\bibliographystyle{IEEEtran}
\bibliography{sigproc}

\begin{thebibliography}{10}
\providecommand{\url}[1]{#1}
\csname url@samestyle\endcsname
\providecommand{\newblock}{\relax}
\providecommand{\bibinfo}[2]{#2}
\providecommand{\BIBentrySTDinterwordspacing}{\spaceskip=0pt\relax}
\providecommand{\BIBentryALTinterwordstretchfactor}{4}
\providecommand{\BIBentryALTinterwordspacing}{\spaceskip=\fontdimen2\font plus
\BIBentryALTinterwordstretchfactor\fontdimen3\font minus
  \fontdimen4\font\relax}
\providecommand{\BIBforeignlanguage}[2]{{%
\expandafter\ifx\csname l@#1\endcsname\relax
\typeout{** WARNING: IEEEtran.bst: No hyphenation pattern has been}%
\typeout{** loaded for the language `#1'. Using the pattern for}%
\typeout{** the default language instead.}%
\else
\language=\csname l@#1\endcsname
\fi
#2}}
\providecommand{\BIBdecl}{\relax}
\BIBdecl

\bibitem{bajaj2014human}
V.~Bajaj and R.~Pachori, ``Human emotion classification from eeg signals using
  multiwavelet transform,'' in \emph{Medical Biometrics, 2014 International
  Conference on}.\hskip 1em plus 0.5em minus 0.4em\relax IEEE, 2014, pp.
  125--130.

\bibitem{zheng2017multichannel}
W.~Zheng, ``Multichannel eeg-based emotion recognition via group sparse
  canonical correlation analysis,'' \emph{IEEE Transactions on Cognitive and
  Developmental Systems}, vol.~9, no.~3, pp. 281--290, 2017.

\bibitem{takahashi2004remarks}
K.~Takahashi \emph{et~al.}, ``Remarks on emotion recognition from bio-potential
  signals,'' in \emph{2nd International conference on autonomous robots and
  agents}, vol.~3, 2004, pp. 1148--1153.

\bibitem{abadi2013multimodal}
M.~K. Abadi, J.~Staiano, A.~Cappelletti, M.~Zancanaro, and N.~Sebe,
  ``Multimodal engagement classification for affective cinema,'' in
  \emph{Affective Computing and Intelligent Interaction (ACII), 2013 Humaine
  Association Conference on}.\hskip 1em plus 0.5em minus 0.4em\relax IEEE,
  2013, pp. 411--416.

\bibitem{scherer1994evidence}
K.~R. Scherer and H.~G. Wallbott, ``Evidence for universality and cultural
  variation of differential emotion response patterning.'' \emph{Journal of
  personality and social psychology}, vol.~66, no.~2, p. 310, 1994.

\bibitem{morioka2015learning}
H.~Morioka, A.~Kanemura, J.-i. Hirayama, M.~Shikauchi, T.~Ogawa, S.~Ikeda,
  M.~Kawanabe, and S.~Ishii, ``Learning a common dictionary for
  subject-transfer decoding with resting calibration,'' \emph{NeuroImage}, vol.
  111, pp. 167--178, 2015.

\bibitem{koch2007gender}
K.~Koch, K.~Pauly, T.~Kellermann, N.~Y. Seiferth, M.~Reske, V.~Backes,
  T.~St{\"o}cker, N.~J. Shah, K.~Amunts, T.~Kircher \emph{et~al.}, ``Gender
  differences in the cognitive control of emotion: An fmri study,''
  \emph{Neuropsychologia}, vol.~45, no.~12, pp. 2744--2754, 2007.

\bibitem{zhu2015cross}
J.-Y. Zhu, W.-L. Zheng, and B.-L. Lu, ``Cross-subject and cross-gender emotion
  classification from eeg,'' in \emph{World Congress on Medical Physics and
  Biomedical Engineering, June 7-12, 2015, Toronto, Canada}.\hskip 1em plus
  0.5em minus 0.4em\relax Springer, 2015, pp. 1188--1191.

\bibitem{soleymani2012multimodal}
M.~Soleymani, J.~Lichtenauer, T.~Pun, and M.~Pantic, ``A multimodal database
  for affect recognition and implicit tagging,'' \emph{IEEE Transactions on
  Affective Computing}, vol.~3, no.~1, pp. 42--55, 2012.

\bibitem{koelstra2012deap}
S.~Koelstra, C.~Muhl, M.~Soleymani, J.-S. Lee, A.~Yazdani, T.~Ebrahimi, T.~Pun,
  A.~Nijholt, and I.~Patras, ``Deap: A database for emotion analysis; using
  physiological signals,'' \emph{IEEE Transactions on Affective Computing},
  vol.~3, no.~1, pp. 18--31, 2012.

\bibitem{lin2017improving}
Y.-P. Lin and T.-P. Jung, ``Improving eeg-based emotion classification using
  conditional transfer learning,'' \emph{Frontiers in human neuroscience},
  vol.~11, p. 334, 2017.

\bibitem{boucenna2014learning}
S.~Boucenna, S.~Anzalone, E.~Tilmont, D.~Cohen, and M.~Chetouani, ``Learning of
  social signatures through imitation game between a robot and a human
  partner,'' \emph{IEEE Transactions on Autonomous Mental Development}, vol.~6,
  no.~3, pp. 213--225, 2014.

\bibitem{dmochowski2012correlated}
J.~P. Dmochowski, P.~Sajda, J.~Dias, and L.~C. Parra, ``Correlated components
  of ongoing eeg point to emotionally laden attention--a possible marker of
  engagement?'' \emph{Frontiers in human neuroscience}, vol.~6, p. 112, 2012.

\bibitem{dmochowski2014audience}
J.~P. Dmochowski, M.~A. Bezdek, B.~P. Abelson, J.~S. Johnson, E.~H. Schumacher,
  and L.~C. Parra, ``Audience preferences are predicted by temporal reliability
  of neural processing,'' \emph{Nature communications}, vol.~5, 2014.

\bibitem{ki2016attention}
J.~J. Ki, S.~P. Kelly, and L.~C. Parra, ``Attention strongly modulates
  reliability of neural responses to naturalistic narrative stimuli,''
  \emph{Journal of Neuroscience}, vol.~36, no.~10, pp. 3092--3101, 2016.

\bibitem{petroni2016age}
A.~Petroni, S.~Cohen, N.~Langer, S.~Henin, T.~Vanderwal, M.~P. Milham, and
  L.~C. Parra, ``Age and sex affect intersubject correlation of eeg throughout
  development,'' \emph{bioRxiv}, p. 089060, 2016.

\bibitem{nummenmaa2012emotions}
L.~Nummenmaa, E.~Glerean, M.~Viinikainen, I.~P. J{\"a}{\"a}skel{\"a}inen,
  R.~Hari, and M.~Sams, ``Emotions promote social interaction by synchronizing
  brain activity across individuals,'' \emph{Proceedings of the National
  Academy of Sciences}, vol. 109, no.~24, pp. 9599--9604, 2012.

\bibitem{blankertz2011single}
B.~Blankertz, S.~Lemm, M.~Treder, S.~Haufe, and K.-R. M{\"u}ller,
  ``Single-trial analysis and classification of erp components - a tutorial,''
  \emph{NeuroImage}, vol.~56, no.~2, pp. 814--825, 2011.

\bibitem{litman2003recognizing}
D.~Litman and K.~Forbes, ``Recognizing emotions from student speech in tutoring
  dialogues,'' in \emph{Automatic Speech Recognition and Understanding, 2003.
  ASRU'03. 2003 IEEE Workshop on}.\hskip 1em plus 0.5em minus 0.4em\relax IEEE,
  2003, pp. 25--30.

\bibitem{martinez2014don}
H.~Martinez, G.~Yannakakis, and J.~Hallam, ``Don't classify ratings of affect;
  rank them!'' \emph{IEEE Transactions on Affective Computing}, vol.~5, no.~3,
  pp. 314--326, 2014.

\bibitem{aigrain2016multimodal}
J.~Aigrain, M.~Spodenkiewicz, S.~Dubuisson, M.~Detyniecki, D.~Cohen, and
  M.~Chetouani, ``Multimodal stress detection from multiple assessments,''
  \emph{IEEE Transactions on Affective Computing}, 2016.

\bibitem{yeh2000more}
A.~Yeh, ``More accurate tests for the statistical significance of result
  differences,'' in \emph{Proceedings of the 18th conference on Computational
  linguistics-Volume 2}.\hskip 1em plus 0.5em minus 0.4em\relax Association for
  Computational Linguistics, 2000, pp. 947--953.

\bibitem{noreen1989computer}
E.~W. Noreen, \emph{Computer-intensive methods for testing hypotheses}.\hskip
  1em plus 0.5em minus 0.4em\relax Wiley New York, 1989.

\end{thebibliography}
\begin{IEEEbiography}[{\includegraphics[width=1in,height=1.25in,clip,keepaspectratio]{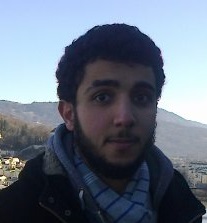}}]{Ayoub Hajlaoui} received the research MS degree in mathematics, computer vision and machine learning from the ENS Cachan engineering school, Paris, 2014, and the MS degree in engineering from the Mines de Nancy (School of Engineering), Nancy, 2014. He is currently working toward the PhD degree in the multi-modal integration, interaction and social signal group, at the Institute of Intelligent Systems and Robotics. The topic of the thesis is the modeling of interactional neurophysiological activity using latent variables. His research interests include machine learning applications in human-computer interaction and affective computing.
\end{IEEEbiography}
\newpage
\begin{IEEEbiography}[{\includegraphics[width=1in,height=1.25in,clip,keepaspectratio]{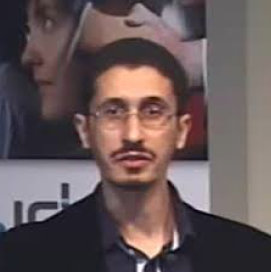}}]{Prof. Mohamed Chetouani} is the head of the IMI2S (Interaction, Multimodal Integration and Social Signal) research group at the Institute for Intelligent Systems and Robotics (CNRS UMR 7222), University Pierre and Marie Curie-Paris 6. He is currently a Full Professor in Signal Processing, Pattern Recognition and Machine Learning at the UPMC. His research activities, carried out at the Institute for Intelligent Systems and Robotics, cover the areas of social signal processing and personal robotics through non-linear signal processing, feature extraction, pattern classification and machine learning. He is also the co-chairman of the French Working Group on Human-Robots/Systems Interaction (GDR Robotique CNRS) and a Deputy Coordinator of the Topic Group on Natural Interaction with Social Robots (euRobotics). He is the Deputy Director of the Laboratory of Excellence SMART Human/Machine/Human Interactions In The Digital Society. In 2016, he was a Visiting Professor at the Human Media Interaction group of University of Twente. He is the coordinator of the ANIMATAS  H2020 Marie Sklodowska Curie European Training Network.
\end{IEEEbiography}
\begin{IEEEbiography}[{\includegraphics[width=1in,height=1.25in,clip,keepaspectratio]{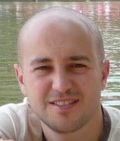}}]{Prof. Slim Essid} is a Full Professor at T\'el\'ecom ParisTech's Department of Images, Data and Signals, and 
the coordinator of the
Audio Data Analysis and Signal Processing team (ADASP, formerly known as the AAO group).
His research interests are in machine learning for temporal data analysis, especially multiview 
learning and structured prediction, with applications in audio, music and multimedia data analysis; 
and multi-modal perception for human behaviour analysis and social signal processing.
He received the state engineering degree from the \'Ecole Nationale d'Ing\'enieurs de Tunis in 2001; the 
M.Sc. (D.E.A.) degree in digital communication systems from the \'Ecole Nationale Sup\'erieure des 
T\'el\'ecommunications, Paris, France, in 2002; the Ph.D. degree from the Universit\'e Pierre et Marie 
Curie (UPMC), in 2005; and the habilitation (HDR) degree from UPMC in 2015.
He has been involved in various French and European research projects among which are Quaero, 
Networks of Excellence Kspace and 3DLife, and collaborative projects REVERIE and LASIE. Over the 
past 10 years, he has collaborated with 12 post-docs and has graduated 6 PhD students; he is 
currently supervising 6 others. He has published over 90 peer-reviewed conference and journal papers 
with more than 50 distinct co-authors. On a regular basis he serves as a reviewer for various 
machine learning, signal processing, audio and multimedia conferences and journals, for instance 
various IEEE transactions, and as an expert for research funding agencies.
\end{IEEEbiography}
\end{document}